\documentclass[preprint,11pt,3p]{elsarticle}

\usepackage{amssymb,amsthm,subfigure,url,amsmath}

\journal{``Nothing can have value without being an object of utility.'' Karl Marx}

\begin{document}

\begin{frontmatter}

\title{On discovering functions in actin filament automata}

\author[label1]{Andrew Adamatzky}
\address[label1]{Unconventional Computing Lab, University of the West of England, Bristol, UK}

\begin{abstract} We simulate an actin filament as an automaton network. Every atom takes two or three states and updates its state, in discrete time, depending on a ratio of its neighbours in some selected state. All atoms/automata simultaneously update their states by the same rule. Two state transition rules are considered. In semi-totalistic Game of Life like actin filament automaton atoms take binary states `0' and `1' and update their states depending on a ratio of neighbours in the state `1'. In excitable actin filament automaton atoms take three states: resting, excited and refractory. A resting atom excites if a ratio of its excited neighbours belong to some specified interval; transitions from excited state to refractory state and from refractory state to resting state are unconditional. In computational experiments we implement mappings of 8-bit input string to 8-bit output string via dynamics of perturbation/excitation on actin filament automata. We assign eight domains in an actin filament as I/O ports. To write {\sc True} to a port we perturb/excite a certain percentage of the nodes in the domain corresponding to the port. We read outputs at the ports after some time interval. A port is considered to be in a state {\sc True} if a number of excited nodes in the port's domain exceed a certain threshold. A range of eight-argument Boolean functions is uncovered in a series of computational trials when all possible configurations of eight-elements binary strings were mapped onto excitation outputs of the I/O domains.
\end{abstract}


\end{frontmatter}

\section{Introduction}

Ideas of information processing on a cytoskeleton network have been proposed by Hameroff and Rasmussen in late 1980s in their designs of tubulin microtubules automata~\cite{hameroff1989information} and a general framework of cytoskeleton automata as sub-cellular information processing networks~\cite{rasmussen1990computational,hameroff1990microtubule}.  Priel, Tuszynski and Horacio development a detailed concept on how information processing could be implemented in actin-tubulin networks of neuron dendrites~\cite{priel2006dendritic}. A signal transmission along the microtubules is implemented via travelling localised patterns of conformation changes or orientations of dipole moments of the tubulin units in tubulin microtubules and ionic waves in actin filaments. A high likehood of existence of travelling localisations (defects, ionic waves, solitons) in tubulin microtubules and actin filaments is supported by a range of independent (bio)-physical models~\cite{tuszynski1995ferroelectric,tuszynski2004results,tuszynski2004ionic, tuszynski2005molecular, tuszynski2005nonlinear, sataric2010solitonic,sataric2011ionic, kavitha2017localized}. A convincing hypothesis is that actin networks in synaptic formations play a role of filtering/processing input information which is further conveyed to and amplified by tubulin microtubules. Thus in present paper we focus on actin filaments. 

Actin is a  protein presented in all eukaryotic cells in forms of globular actin (G-actin) and filamentous actin (F-actin)~\cite{straub1943actin, korn1982actin, szent2004early}. G-actin, polymerises in double helix of filamentous actin, during polymerisation G-actin units slightly change their shapes and thus become F-actin units~\cite{oda2009nature}.   The actin networks play a key role in information processing~\cite{fifkova1982cytoplasmic, kim1999role, dillon2005actin, cingolani2008actin} in living cells. Previously we have demonstrated how to implement Boolean, multi-valued and quantum logical gates on coarse-grained models of actin filaments using cellular automata, quantum automata and a lattice with Morse potential approaches~\cite{siccardi2015actin, siccardi2016boolean, siccardi2016quantum, siccardi2016logical, siccardi2017models}.  Theoretical designs of actin-based logical circuits realise logical gates via collisions between travelling localisations. Such an approach assumes that we can address nearly every atom in the actin molecule~\cite{adamatzky2017logical} or control exact timing of the collisions between travelling localisations~\cite{siccardi2017models}. Such assumptions might prove to be unrealistic when experimental laboratory implementations concern.  This is why, in present paper, we consider a less restrictive, than in previous implementations, way of executing computation on protein polymer: to probe, relatively large, portions of an actin filament as I/O and uncover Boolean functions implemented via input to output mapping.  The approach proposed is novel and have not been considered before. Another original feature of the presented results is that we employ a detailed model of several actin units arranged in the helix. The model is introduced in Sect.~\ref{model}. To discover Boolean functions implementable in the actin filament we split the helix into eight domains. We perturb the domains in all possible combinations of excitation representing state of 8-bit strings and record their outputs. A mapping between an input and output sets of binary strings is constructed then. This is show in Sect.~\ref{miningFunction}. We discuss limitations of the approach and future developments in Sect.~\ref{discussion}.

\section{Actin filament automata}
\label{model}

\begin{figure}[!tbp]
\centering
\includegraphics[width=1.\textwidth]{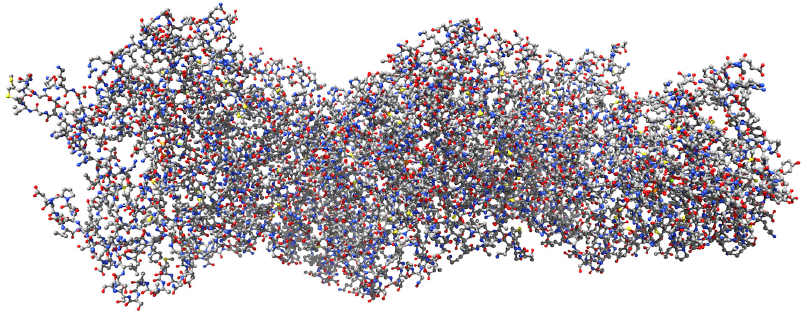}
\label{fig:actinchain}
\caption{Pseudo-atomic model of F-actin~\cite{galkin2015near} in  Corey-Pauling-Kolun colouring.}
\end{figure}

\begin{figure}[!tbp]
    \centering
    \includegraphics[width=\textwidth]{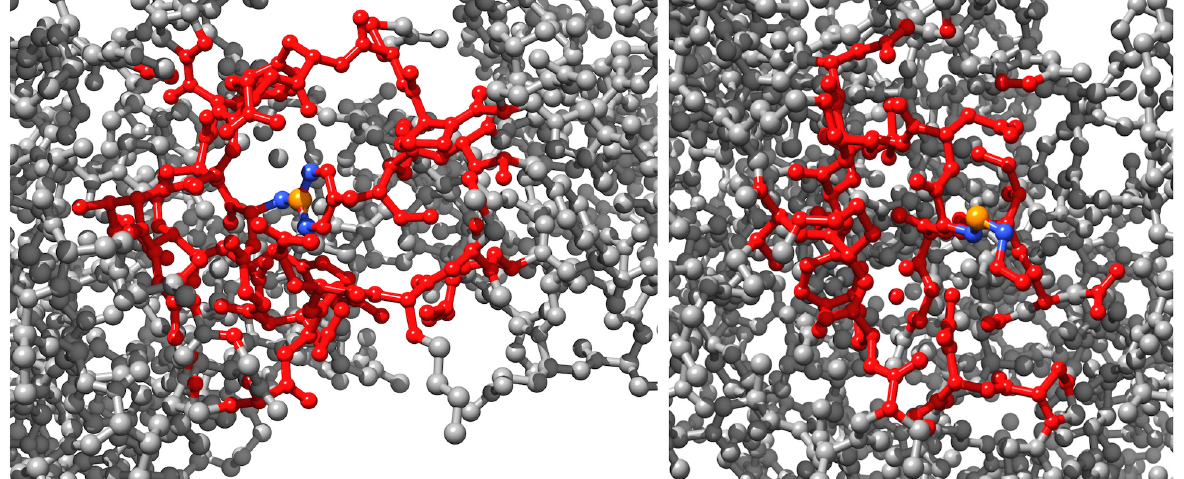}
    \caption{Examples of neighbourhoods. Central nodes, `owners' of the neighbourhoods are coloured orange, their hard neighbours are blue and their soft neighbours are red.}
    \label{fig:neighbourhoods}
\end{figure}

We employed a pseudo-atomic model of F-actin filament (Fig.~\ref{fig:actinchain}) reconstructed by Galkin et al.~\cite{galkin2015near} at 4.7~{\AA} resolution using direct electron detector, cryoelectron microscopy and the forces imposed on actin filaments in thin films~\footnote{PDB file can be downloaded here \url{https://www.rcsb.org/structure/3J8I}}. The model has 14800 atoms and is composed of six F-actin molecules. Following our previous convention~\cite{adamatzky2017dynamics} we represent F-actin filament as a graph $\mathcal{F} = \langle \bf{V}, \bf{E}, \bf{C}, \bf{Q}, f \rangle$,  where $\bf{V}$ is a set of nodes, $\bf{E}$ is a set of edges, $\bf{C}$ is a set of Euclidean coordinates of nodes form $\bf{V}$,  $\bf{Q}$ is a set of node states, $f: \bf{Q} \times [0,1] \rightarrow \bf{Q}$ is a node state transition function. Each atom from pseudo-atomic model of F-actin filament is represented by a node from $\bf{V}$ with its 3D coordinates being a member of $\bf{C}$; atomic bonds are represented by $\bf{E}$. Each node $p \in \bf{V}$ takes states from a finite set $\bf{Q}$. All nodes updated their states simultaneously in discrete time. A node $p$ updates its state depending on its current state $p^t$ and ratio $\gamma (p)^t$ of its neighbours being in some selected state $\star$. We consider two types of a node neighbourhood. Let $u(p)$ be nodes from $\bf{V}$ that are connected with an edge with a node $p$, they correspond to atoms connected by chemical bonds with atom $p$. We call them hard-neighbours because their neighbourhood is determined by chemical structure of F-actin. A ratio of nodes with one hard neighbour is 0.298, two hard neighbours 0.360, three hard neighbours 0.341, and four neighbours 0.001.

\begin{figure}[!tbp]
    \centering
   \includegraphics[width=0.7\textwidth]{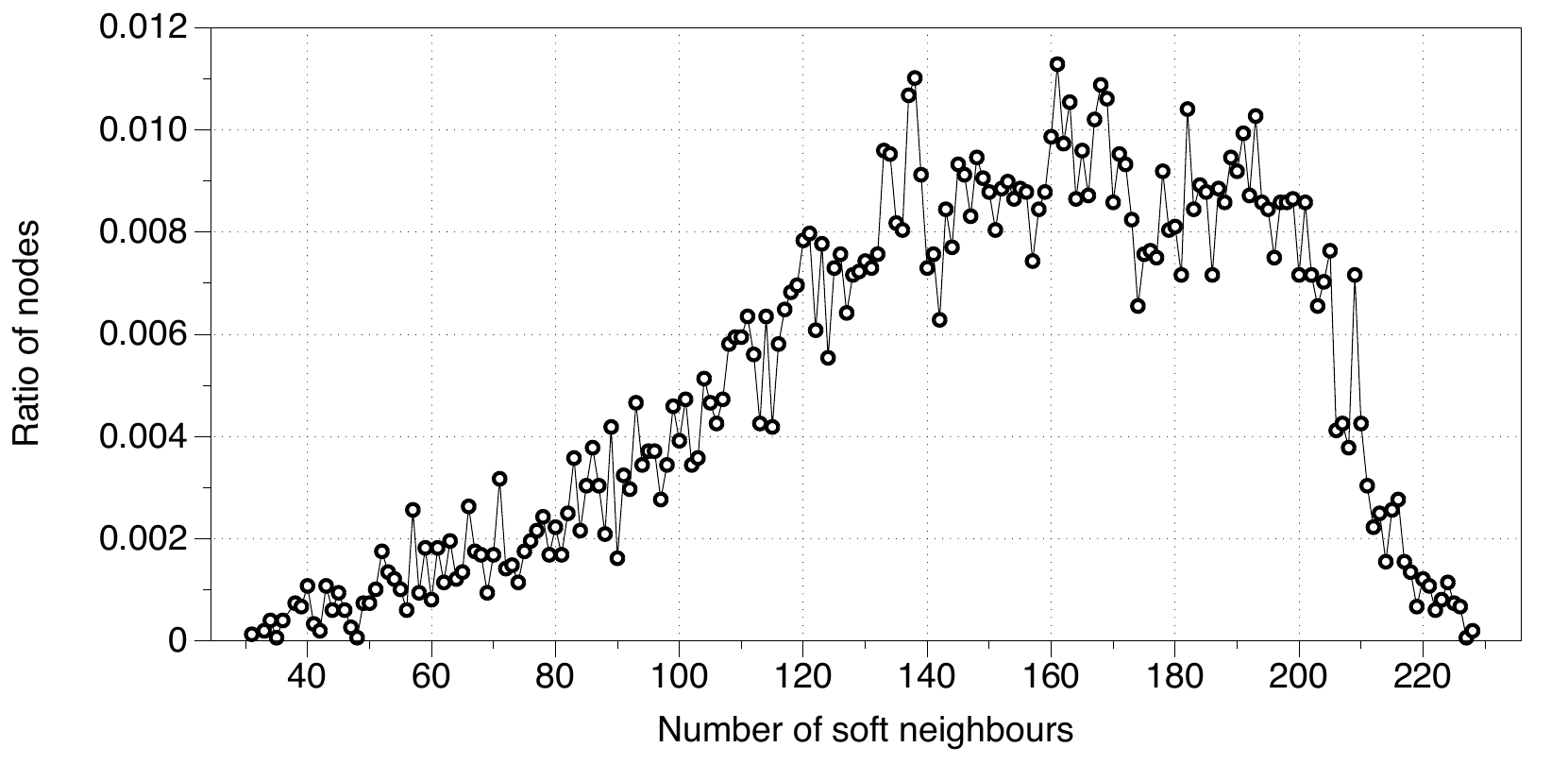}
    \caption{Distribution of a ratio of nodes vs. numbers of their soft neighbours, $\rho=10$.}
    \label{fig:distsoftneighbours}
\end{figure}

Actin molecule is folded in 3D. Let $\delta$ be an average distance between two hard-neighbours, for F-actin $\delta=1.43${\AA} units. Let $w(p)$ be nodes of $\mathcal F$ that are at distance not exceed $\rho$, in the Euclidean space, from node $p$. We call them soft neighbours because their neighbourhood is determined by 3D structure of F-actin.
Thus, each node $p$ has two neighbourhoods: hard neighbourhood $u(p) = \{ s \in {\bf V}: (ps) \in {\bf E} \}$ (actin automata with hard neighbourhood were firstly proposed by us in \cite{adamatzky2017dynamics}),
and  soft neighbourhood $w(p) = \{ s \in {\bf V}: s \notin u(p) \text{ and } d(c_p, c_s) \leq \rho \}$, where $d(c_p, c_s)$ is a distance between nodes $p$ and $s$ in 3D Euclidean space and $c_s, c_p \in \bf{C}$. We have chosen $\rho=10$~\AA, which is seven times more than an average Euclidean distance 1.42~\AA\, between two hard-neighbours. Examples of neighbourhoods are shown in Fig.~\ref{fig:neighbourhoods}.   Distribution of a number of soft neighbours versus a ratio of nodes with such number of soft neighbours is shown in Fig.~\ref{fig:distsoftneighbours};  nearly half of the nodes (ratio 0.45) has from 133 to 185 neighbours. The ratio $\gamma(p)^t$  is calculated as $\gamma(p)^t=\frac{|{s \in u(p): s^t=\star}|+ \mu \cdot |{s \in w(p): s^t=\star}|}{|u(p)|+|w(p)}|$, where $|\mathbf{S}|$ is a number of elements in the set $\mathbf{S}$; we used $\mu=0.9$ in experiments reported.

We consider two species of family $\mathcal{F}$: semi-totalistic automaton $\mathcal{G}=\langle \mathbf{V}, \mathbf{E}, \mathbf{C}, \{\star, \circ \}, f^{G} \rangle$ and excitable automaton $\mathcal{E}=\langle \mathbf{V}, \mathbf{E}, \mathbf{C}, \{ \star, \circ, \bullet \},  f^{E} \rangle$.  The rules $f^{G}$ and $f^{E}$ are defined as follows:
\begin{equation}
p^{t+1}=f^{G}(p)=
\begin{cases}
\star, \text{ if } ((p^t=\circ)\land(\theta'_\circ \leq \gamma(p)^t \leq \theta''_\circ))\lor ((p^t=\star)\land(\theta'_\star \leq \gamma(p)^t \leq \theta''_\star))\\
\circ, \text{ otherwise }
\end{cases}
\label{eqG}
\end{equation}
\begin{equation}
p^{t+1}=f^{E}(p)=
\begin{cases}
\star, \text{ if } ((p^t=\circ)\land(\theta'_\circ \leq \gamma(p)^t \leq \theta''_\circ))\\
\bullet, \text{ if }  p^t=\circ\\
\circ, \text{ otherwise }
\circ, \text{ otherwise }
\end{cases}
\label{eqE}
\end{equation}

We chosen intervals $[\theta'_\circ,\theta''_\circ]=[\theta'_\star,\theta''_\star]=[0.25,0.375]$ for $\mathcal{G}$ and $[\theta'_\circ,\theta''_\circ]=[0.15,0.25]$ for $\mathcal{E}$ because they support localised modes of excitation, i.e. a perturbation of the automata at a single site or a compact domain of several sties does not lead to an excitation spreading all over the automaton. Localised excitations emerged at different input domains can interact with other and the results of their interactions in the output domains will represent values of a logical function computed.   

\begin{figure}[!tbp]
    \centering
  \includegraphics[]{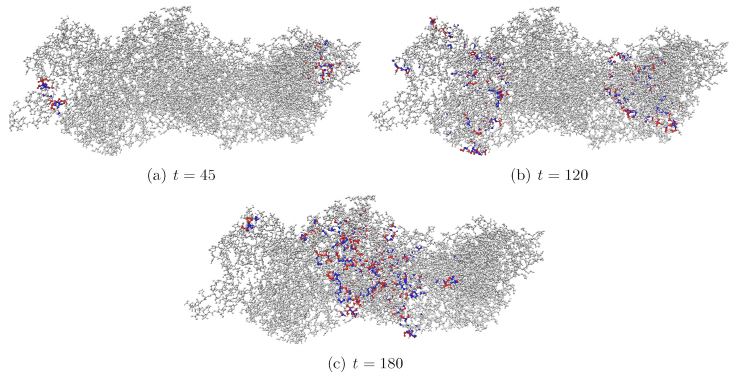}
    \caption{Annihilation of excitation wave-fronts in $\mathcal{E}$ for $[\theta'_\circ,\theta''_\circ]=[0.125,1]$.}
    \label{fig:twowaves}
\end{figure}

Automaton $\mathcal{G}$ is a Game of Life like automaton~\cite{conway1970game,adamatzky2010game}. Speaking in the Game of Life lingo we can say that a dead node $\circ$ becomes alive $\star$ if a ratio of live nodes in its neighbourhood lies inside interval $[\theta'_\circ,\theta''_\circ]$; a live node $\star$ remains alive if a ratio of live nodes in its neighbourhood lies inside interval $[\theta'_\star,\theta''_\star]$. 
Automaton $\mathcal{E}$ is a Greenberg-Hastings~\cite{greenberg1978spatial} like automaton:  a resting node $\circ$ excites if a ratio of excited nodes in its neighbourhood lies inside interval $[\theta'_\circ,\theta''_\circ]$; and excited node $\star$ takes refractory state $\bullet$ in next step of development, and a refractory $\bullet$ returns to resting state $\circ$.
Rules of Conway's Game of Life could be interpreted as Eq. (\ref{eqG}) have perturbation intervals $[\theta'_\circ,\theta''_\circ]=[0.375,0.375]$ and $[\theta'_\star,\theta''_\star]=[0.25,0.375]$,  of Greenberg-Hasting automata in terms of Eq. (\ref{eqE}) have interval $[\theta'_\circ,\theta''_\circ]=[0.125,1]$. The exact intervals of perturbation for the Game of Life and the Greenberg-Hasting automata are proven to be not useful for mining functions. This is because $\mathcal G$ with the Game of Life interval does not show any sustainable dynamics of excitation, and  $\mathcal E$ with Greenberg-Hasting interval exhibits `classical' waves of excitation, where two colliding waves annihilate (Fig.~\ref{fig:twowaves}).  

The model was implemented in Processing. Data are analysed in Matlab. Patterns of excitation dynamics are visualised in Processing and Chimera.

\section{Discovering functions}
\label{miningFunction}

\begin{figure}[!tbp]
    \centering
     \includegraphics[width=\textwidth]{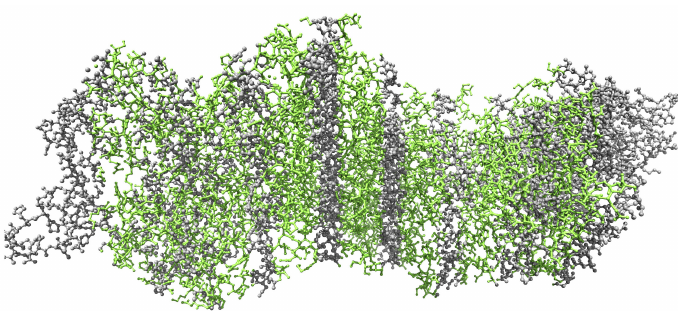}
    \caption{Nodes of I/O domains $D_0 \ldots D_7$ are shown by green colour.}
    \label{fig:IOSites}
\end{figure}

\begin{figure}[!tbp]
\includegraphics[]{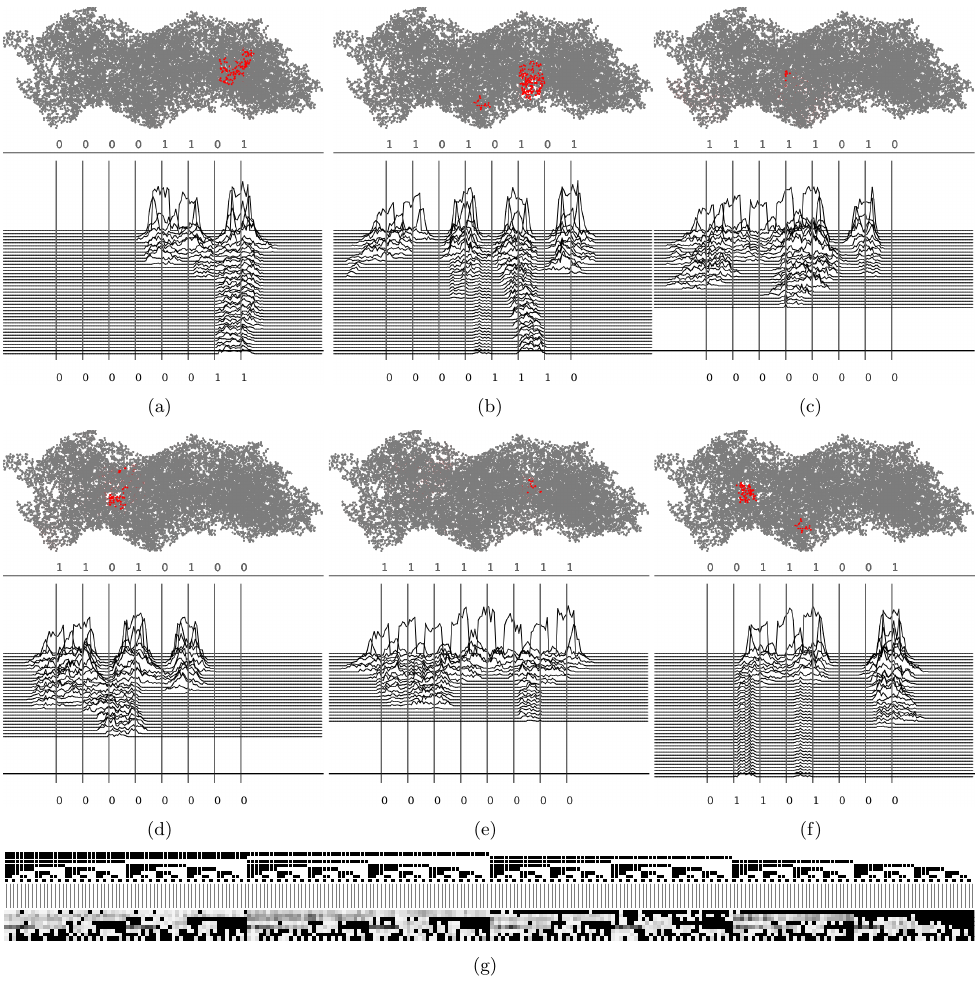}
    \caption{Discovering Boolean functions in automaton $\mathcal{G}$. (a--f) Examples of excitation dynamics in automaton $\mathcal{G}$, $\theta'_\circ=\theta'_\star=0.25$ and $\theta''_\circ=\theta''_\star=0.375$. Projection of actin filament on $z$ plane is shown in grey; projection of nodes being in state $\star$  by the moment of recording inputs are shown in red. Plots show values of activity, i.e. a number of nodes in state $\star$ along $x$ coordinate. See videos of experiments in \protect\url{https://doi.org/10.5281/zenodo.1312141}.
    (g)~Visualisation of register mapping implemented by automaton $\mathcal{G}$.}
    \label{fig:exampleDynamics}
\end{figure}

\begin{figure}
    \centering
     \includegraphics[width=0.9\textwidth]{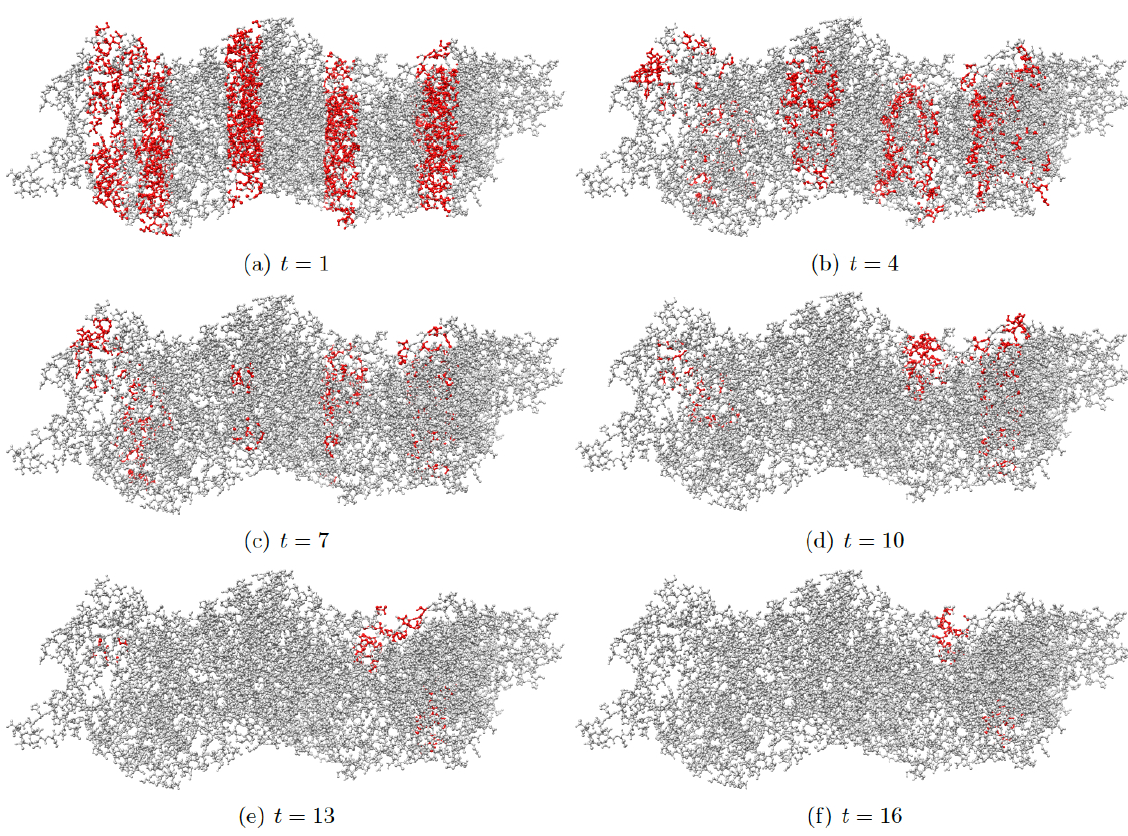}
    \caption{Snapshots of excitation dynamics of automaton $\mathcal G$ in response to the input 11010101. See videos of experiments in \protect\url{https://doi.org/10.5281/zenodo.1312141}.}
    \label{fig:my_label}
\end{figure}

\begin{table}[!tbp]
    \caption{Fragment of experimentally obtained mapping $\bf{S}$ to $\bf{W}$ for automaton $\mathcal{G}$.}
    \centering
    \begin{footnotesize}
    \begin{tabular}{c|cccccccc}
$(I_0I_1I_2I_3I_4I_5I_6I_7)$	&	$w_0$	&	$w_1$	&	$w_2$	&	$w_3$	&	$w_4$	&	$w_5$	&	$w_6$	&	$w_7$	\\ \hline
1011100	&	0	&	0	&	0.01	&	0.01	&	0.14	&	0.03	&	0.01	&	0	\\
1011101	&	0.01	&	0.03	&	0.01	&	0.01	&	0.2	&	0.03	&	0.01	&	0.03	\\
1011110	&	0	&	0	&	0	&	0	&	0.14	&	0.02	&	0.01	&	0.01	\\
1011111	&	0	&	0.01	&	0.01	&	0.02	&	0.25	&	0.04	&	0.02	&	0.02	\\
1100000	&	0	&	0.04	&	0.04	&	0	&	0	&	0	&	0	&	0	\\
1100001	&	0	&	0.02	&	0.02	&	0	&	0	&	0	&	0	&	0.01	\\
1100010	&	0	&	0.03	&	0.05	&	0	&	0	&	0.01	&	0.02	&	0	\\
1100011	&	0	&	0.05	&	0.03	&	0	&	0	&	0	&	0	&	0.02	\\
1100100	&	0	&	0.06	&	0.04	&	0	&	0	&	0.02	&	0	&	0	\\
1100101	&	0.01	&	0.06	&	0.04	&	0	&	0	&	0.04	&	0.03	&	0.02	\\
    \end{tabular}
    \end{footnotesize}
    \label{tab:examplemapping}
\end{table}

We encode Boolean values `0' ({\sc False}) and `1' ({\sc True}) in perturbations of selected domains $\mathbf{D}$ and extract a range of mappings $\{0, 1\}^m \rightarrow \{0, 1\}^m$, $m \in \mathbf{N}$, implementable by the actin filament automaton. Assume input and output tuples  $\mathbf{I} \in \{0, 1\}^m$ and $\mathbf{O} \in \{0, 1\}^m$,$m=8$, the actin automaton implements $\mathbf{I} \rightarrow \mathbf{D} \rightarrow \mathbf{O}$.
We implement computation on actin filament automaton as follows. Eight cylinders across $(xy)$-plane with coordinates 
$D_i = \{ p \in \mathcal{V}: \text{abs}(p_x-k(i)) < r_s \}$, $0 \leq i < 8$,
$k(i)=15\cdot(i+1)$, are assigned as input-output domains (Fig.~\ref{fig:IOSites}). These are mapped onto Boolean inputs $\mathbf{I}=(I_0, \ldots, I_8)$ and outputs $\mathbf{O}=(O_0, \ldots, O_8)$ as follows: $I_z=1$ if $\sum_{p^0 \in D_z} > \kappa$, otherwise $I_z=0$, and $O_z=1$ if $\sum_{p^\zeta \in D_z} > \kappa$, otherwise $O_z=0$,  in present paper we chosen $\kappa=0$ and $\zeta=40$. 

Domains from $\mathbf{D}$ at time step $t=0$ are excited with probability $p$ determined by values of inputs $\mathbf{I}$:
if a node $p$ belongs to $\bf{D}_i$ and $s_i=1$ the node takes state $\star$ at the beginning of evolution, $p^0=\star$ with probability $p$. We read outputs after $\zeta=40$ steps of automaton evolution. As soon as 40 iterations occurred ($t=41$) we measure states of nodes in the domains $\bf{D}_i$, $s_i \in \{0, 1 \}$, and assign outputs depending on the excitation:  $O_i=1$ if $|\{ p \in \bf{D}: p^t=\star \}|>\kappa$, $\kappa=0$. Stimulation runs for $h$ trials (repeated simulation of automaton) with all possible configurations of $\mathbf{I}$, $h=100$, where frequencies of outputs are calculated as $W_i=w_i+I^T_i$, $0 \leq i < 8$ where $T$ is a trial number, $T=1, \ldots, h$. By the end of the experiments we normalise $\bf{W}$ as $w_i=w_i/h$, $h$ is a number of trials.

\begin{figure}[!tbp]
    \centering 
    \includegraphics[]{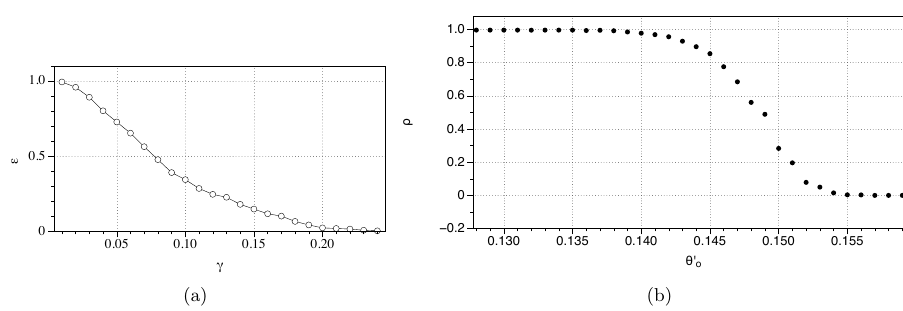} 
    \caption{
    (b)~Ratio $\epsilon$ of transitions where at least one entry in $\bf{W}$ exceeds $\gamma$.
    (c)~Dependence of the ratio $\rho$ of outputs in state 1 to an overall number of outputs of the lower threshold of excitation $\theta'_\circ$; upper threshold $\theta''_\circ=0.25$ was kept constant. }
    \label{fig:thresholdplots}
\end{figure}

Examples of perturbation dynamics of automaton $\mathcal{G}$ for various input sequences are shown in Fig.~\ref{fig:exampleDynamics}af. Example of a fragment of  $\bf{W}$ obtained in 100 trials with automaton $\bf{G}$ is shown in Tab.~\ref{tab:examplemapping}. Visualisation of mapping $\bf{S} \rightarrow \bf{W}$ is presented in Fig.~\ref{fig:exampleDynamics}g. There lexicograpically ordered elements of $\bf{S}$ are shown by black (`1') and white (`0') squares: top row from (0000000) on the right to (11111111) on the left. Corresponding elements of $\bf{W}$ are by shown by gradations of grey $255 \cdot w_i$. From $\bf{W}$ we extract values of outputs $\bf{O}$ for various ranges of $\gamma \in [0,1]$ as follows: $O_i=1$ if $w_i>\gamma$, and $O_i=0$ otherwise.

\begin{table}[!tbp]
 \caption{Functions implemented by  (a)~$\mathcal{G}$ automatom, $\theta'_\circ=\theta'_\star=0.25$ and $\theta''_\circ=\theta''_\star=0.375$, 
 and (b)~$\mathcal{E}$ automaton, $\theta'_\circ=0.15$ and $\theta''_\circ=0.25$. 
 for various values of reliability threshold $\gamma$.}
    \centering
    \begin{footnotesize}
    \subfigure[]{
    \begin{tabular}{p{0.6cm}|p{12cm}}
    $\gamma$ &  Functions  \\ \hline
    0.15 & $O_1=I_0 \cdot \overline{I_1 } \cdot I_2 \cdot \overline{I_3 } \cdot I_4 \cdot I_5 \cdot I_6 \cdot I_7 
$; \newline
$O_2=\overline{I_0 } \cdot \overline{I_1 } \cdot I_2 \cdot I_3 \cdot I_7 \cdot (I_4 \cdot I_5 \cdot \overline{I_6 } + I_4 \cdot \overline{I_5 } \cdot I_6 + \overline{I_4 } \cdot I_5 \cdot I_6) 
$; \newline
    $O_4=\overline{I_0 } \cdot \overline{I_1 } \cdot \overline{I_2 } \cdot I_3 \cdot I_4 \cdot (\overline{I_5 } \cdot \overline{I_7 } + I_6 \cdot I_7 + I_5 \cdot I_6 \cdot \overline{I_7}) $\\
    0.2 & $O_4=I_3 \cdot I_4 \cdot (\overline{I_0 } \cdot I_1 \cdot \overline{I_2 } \cdot I_5 \cdot I_7 + \overline{I_0 } \cdot \overline{I_2 } \cdot \overline{I_5 } \cdot \overline{I_6 } \cdot \overline{I_7 } + I_0 \cdot \overline{I_1 } \cdot \overline{I_2 } \cdot \overline{I_5 } \cdot I_6 + \overline{I_0 } \cdot I_1 \cdot I_2 \cdot \overline{I_5 } \cdot I_6 \cdot \overline{I_7 } + I_0 \cdot I_1 \cdot I_2 \cdot \overline{I_5 } \cdot I_6 \cdot I_7 + I_0 \cdot \overline{I_1 } \cdot \overline{I_2 } \cdot \overline{I_5 } \cdot \overline{I_6 } \cdot I_7 + \overline{I_0 } \cdot \overline{I_1 } \cdot \overline{I_2 } \cdot I_5 \cdot I_6 \cdot \overline{I_7 } + \overline{I_0 } \cdot \overline{I_1 } \cdot \overline{I_2 } \cdot \overline{I_5 } \cdot I_6 \cdot I_7)$\\
    0.22 & $O_4=\overline{I_2 } \cdot I_3 \cdot I_4 \cdot (\overline{I_0 } \cdot I_1 \cdot I_5 \cdot I_6 \cdot I_7 + \overline{I_0 } \cdot \overline{I_1 } \cdot \overline{I_5 } \cdot \overline{I_6 } \cdot \overline{I_7 } + I_0 \cdot \overline{I_1 } \cdot \overline{I_5 } \cdot \overline{I_6 } \cdot I_7 + \overline{I_0 } \cdot \overline{I_1 } \cdot I_5 \cdot I_6 \cdot \overline{I_7 } + \overline{I_0 } \cdot \overline{I_1 } \cdot \overline{I_5 } \cdot I_6 \cdot I_7) $\\
    0.23 & $O_4=\overline{I_0 } \cdot \overline{I_2 } \cdot I_3 \cdot I_4 \cdot (\overline{I_1 } \cdot I_5 \cdot I_6 \cdot 
    \overline{I_7} + \overline{I_1 } \cdot \overline{I_5 } \cdot I_6 \cdot I_7 + I_1 \cdot I_5 \cdot I_6 \cdot I_7 + \overline{I_1} \cdot \overline{I_5} \cdot  \overline{I_6} \cdot \overline{I_7}) $\\
    0.24 & $O_4=\overline{I_0 } \cdot \overline{I_2 } \cdot I_3 \cdot I_4 \cdot I_6 \cdot I_7 \cdot (\overline{I_1 } \cdot \overline{I_5 } + I_1 \cdot I_5) $\\
    0.25     &  $O_4=\overline{I_0}\cdot I_1\cdot \overline{I_2}\cdot I_3  \cdot I_4  \cdot I_5  \cdot I_6  \cdot I_7$ \\
    \end{tabular}
    }\label{tab:G}
   \subfigure[]{
   \begin{tabular}{p{0.6cm}|p{12cm}}
    $\gamma$ &  Functions  \\ \hline
    0.7 &
    $O_0=\overline{I_0 } \cdot I_1 \cdot \overline{I_2 } \cdot I_7 \cdot (\overline{I_3 } \cdot \overline{I_4 } \cdot I_5 \cdot \overline{I_6 } + I_3 \cdot I_4 \cdot \overline{I_5 } \cdot I_6) $ \newline
    $O_1=I_0 \cdot \overline{I_1 } \cdot I_2 \cdot \overline{I_3 } \cdot I_4 + \overline{I_0 } \cdot I_1 \cdot \overline{I_2 } \cdot I_3 \cdot I_4 \cdot \overline{I_5 } \cdot I_7 + \overline{I_0 } \cdot I_1 \cdot \overline{I_2 } \cdot I_3 \cdot \overline{I_4 } \cdot I_5 \cdot \overline{I_6 } + \overline{I_0 } \cdot I_1 \cdot \overline{I_2 } \cdot I_3 \cdot \overline{I_4 } \cdot I_6 \cdot \overline{I_7 } + \overline{I_0 } \cdot I_1 \cdot I_2 \cdot \overline{I_3 } \cdot I_4 \cdot \overline{I_5 } \cdot \overline{I_6 } \cdot I_7 + \overline{I_0 } \cdot I_1 \cdot I_2 \cdot \overline{I_3 } \cdot I_4 \cdot I_5 \cdot I_6 \cdot I_7 + I_0 \cdot \overline{I_1 } \cdot I_2 \cdot I_3 \cdot \overline{I_4 } \cdot I_5 \cdot \overline{I_6 } \cdot I_7 + \overline{I_0 } \cdot I_1 \cdot \overline{I_2 } \cdot I_3 \cdot \overline{I_4 } \cdot I_5 \cdot I_6 \cdot I_7 $ \newline
    $O_2=I_0 \cdot \overline{I_1 } \cdot I_2 \cdot \overline{I_3 } \cdot I_4 + \overline{I_0 } \cdot I_1 \cdot I_2 \cdot \overline{I_3 } \cdot I_4 \cdot \overline{I_5 } + I_0 \cdot \overline{I_1 } \cdot I_2 \cdot I_3 \cdot \overline{I_4 } \cdot I_5 \cdot \overline{I_7 } + \overline{I_0 } \cdot I_1 \cdot \overline{I_2 } \cdot I_3 \cdot \overline{I_4 } \cdot I_6 \cdot \overline{I_7 } + \overline{I_0 } \cdot I_1 \cdot \overline{I_2 } \cdot I_3 \cdot \overline{I_4 } \cdot \overline{I_6 } \cdot I_7 + \overline{I_0 } \cdot I_1 \cdot I_3 \cdot \overline{I_4 } \cdot I_5 \cdot \overline{I_6 } \cdot \overline{I_7 } + \overline{I_0 } \cdot I_1 \cdot \overline{I_2 } \cdot I_3 \cdot I_4 \cdot \overline{I_5 } \cdot \overline{I_6 } \cdot I_7 + I_0 \cdot I_1 \cdot \overline{I_2 } \cdot I_3 \cdot \overline{I_4 } \cdot I_5 \cdot \overline{I_6 } \cdot I_7 + I_0 \cdot \overline{I_1 } \cdot I_2 \cdot I_3 \cdot \overline{I_4 } \cdot I_5 \cdot \overline{I_6 } \cdot I_7 + \overline{I_0 } \cdot I_1 \cdot I_2 \cdot I_3 \cdot \overline{I_4 } \cdot I_5 \cdot \overline{I_6 } \cdot I_7 + \overline{I_0 } \cdot I_1 \cdot I_2 \cdot \overline{I_3 } \cdot I_4 \cdot I_5 \cdot I_6 \cdot \overline{I_7 } + \overline{I_0 } \cdot I_1 \cdot \overline{I_2 } \cdot I_3 \cdot \overline{I_4 } \cdot I_5 \cdot I_6 \cdot I_7 $ \newline
    $O_3=I_2 \cdot \overline{I_3 } \cdot I_4 \cdot \overline{I_5 } \cdot \overline{I_6 } \cdot I_7 + I_0 \cdot I_3 \cdot \overline{I_4 } \cdot I_5 \cdot \overline{I_6 } \cdot I_7 + \overline{I_0 } \cdot I_1 \cdot I_2 \cdot I_3 \cdot \overline{I_4 } \cdot I_5 \cdot \overline{I_7 } + \overline{I_0 } \cdot I_1 \cdot I_3 \cdot \overline{I_4 } \cdot I_5 \cdot \overline{I_6 } \cdot I_7 + \overline{I_1 } \cdot I_2 \cdot I_3 \cdot \overline{I_4 } \cdot I_5 \cdot I_6 \cdot \overline{I_7 } + I_0 \cdot \overline{I_1 } \cdot I_2 \cdot \overline{I_3 } \cdot I_4 \cdot \overline{I_5 } \cdot \overline{I_7 } + \overline{I_0 } \cdot I_1 \cdot I_2 \cdot \overline{I_3 } \cdot I_4 \cdot \overline{I_5 } \cdot \overline{I_7 } + \overline{I_0 } \cdot \overline{I_2 } \cdot I_3 \cdot \overline{I_4 } \cdot I_5 \cdot I_6 \cdot \overline{I_7 } + I_1 \cdot \overline{I_2 } \cdot I_3 \cdot \overline{I_4 } \cdot I_5 \cdot \overline{I_6 } \cdot \overline{I_7 } + I_0 \cdot I_1 \cdot \overline{I_2 } \cdot I_3 \cdot \overline{I_4 } \cdot I_5 \cdot I_6 + I_0 \cdot \overline{I_1 } \cdot I_2 \cdot I_3 \cdot \overline{I_4 } \cdot I_5 \cdot \overline{I_6 } \cdot \overline{I_7 } + \overline{I_0 } \cdot \overline{I_1 } \cdot I_2 \cdot I_3 \cdot \overline{I_4 } \cdot I_5 \cdot \overline{I_6 } \cdot I_7 + I_0 \cdot I_1 \cdot I_2 \cdot \overline{I_3 } \cdot I_4 \cdot \overline{I_5 } \cdot I_6 \cdot \overline{I_7 } + I_0 \cdot \overline{I_1 } \cdot I_2 \cdot \overline{I_3 } \cdot I_4 \cdot \overline{I_5 } \cdot I_6 \cdot I_7 + \overline{I_0 } \cdot I_1 \cdot I_2 \cdot \overline{I_3 } \cdot I_4 \cdot \overline{I_5 } \cdot I_6 \cdot I_7 + \overline{I_0 } \cdot \overline{I_1 } \cdot I_2 \cdot \overline{I_3 } \cdot I_4 \cdot \overline{I_5 } \cdot \overline{I_6 } \cdot \overline{I_7} $ \newline
    Functions realises on outputs $O_4$ to $O_7$ are not shown.
    \\
    0.8 & 
    $O_2=I_3 \cdot \overline{I_4 } \cdot I_5 \cdot I_7 \cdot (\overline{I_0 } \cdot I_1 \cdot \overline{I_2 } + I_0 \cdot \overline{I_1 } \cdot I_2 \cdot \overline{I_6}) $\newline 
    $O_3=\overline{I_6 } \cdot (I_0 \cdot \overline{I_1 } \cdot I_3 \cdot \overline{I_4 } \cdot I_5 \cdot I_7 + \overline{I_0 } \cdot I_1 \cdot I_3 \cdot \overline{I_4 } \cdot I_5 \cdot I_7 + I_0 \cdot \overline{I_1 } \cdot I_2 \cdot I_3 \cdot \overline{I_4 } \cdot I_5 \cdot \overline{I_7 } + I_0 \cdot \overline{I_1 } \cdot I_2 \cdot \overline{I_3 } \cdot I_4 \cdot \overline{I_5 } \cdot I_7 + I_0 \cdot I_1 \cdot \overline{I_2 } \cdot I_3 \cdot \overline{I_4 } \cdot I_5 \cdot I_7) $ \newline
    $O_4=\overline{I_6 } \cdot (I_2 \cdot I_3 \cdot \overline{I_4 } \cdot I_5 \cdot I_7 + \overline{I_1 } \cdot I_2 \cdot \overline{I_3 } \cdot I_4 \cdot \overline{I_5 } \cdot I_7 + I_0 \cdot I_1 \cdot \overline{I_2 } \cdot I_3 \cdot \overline{I_4 } \cdot I_5 + I_0 \cdot \overline{I_1 } \cdot \overline{I_2 } \cdot I_3 \cdot \overline{I_4 } \cdot I_5 \cdot I_7) $ \newline
    $O_5=\overline{I_2 } \cdot I_3 \cdot \overline{I_4 } \cdot I_5 \cdot \overline{I_6 } \cdot I_7 \cdot (I_0 + \overline{I_1}) $\\
    0.9 & 
$O_5=\overline{I_1 } \cdot I_3 \cdot \overline{I_4 } \cdot I_5 \cdot \overline{I_6 } \cdot I_7 \cdot (I_0 +\overline{I_2})$\\ 
   \end{tabular}
   }\label{tab:functions}
   \end{footnotesize}
\end{table}

Boolean functions, in the form $O_i=f(I_0 \ldots I_7)$, realisable by automata $\mathcal{G}$ and $\mathcal{E}$ are listed in Tab.~\ref{tab:functions}. In automaton $G$ a ratio $\epsilon$ of I/O transitions where at least one element of $\bf{W}$ exceeds $\gamma$ shows quadratic decrease with increase of $\gamma$ (Fig.~\ref{fig:thresholdplots}a); the same applies to automaton $\bf{E}$. This reflects both a decrease in a number of functions realisable on output domains and a decrease of the functions complexity in terms of the arguments.  A number of functions implementable in $\mathcal{F}$ polynomially decrease with increase of $\theta'_\circ$ (Fig.~\ref{fig:thresholdplots}b).

\section{Discussion}
\label{discussion}

We demonstrated implementation of logical functions on automaton models of actin filaments. The approach was inspired by the `evolution in materio' framework~\cite{harding2005evolution,miller2014evolution,harding2016discovering} on implementing computation without knowing exact physical structure of a computing substrate. Propagating patterns in the automaton $\mathcal{G}$ can be see as discrete analogies of vibration excitation~\cite{davydov1979solitons,sirenko1996dynamics,pokorny1997vibrations,pokorny2004excitation}. The dynamics of automaton $\mathcal{E}$ is a finite-state machine analog of the ionic waves, theoretical models of which are well studied in a context of tubulin microtubules and actin filaments~\cite{sataric2009actin,tuszynski2004ionic,sekulic2011nonlinear, priel2008nonlinear,sataric2010solitonic}.  How feasible is the approach? So far there are no experimental data on vibration modes of a single strand, or even a bundle of actin filaments or tubulin tubes,  of a cytoskeleton polymer~\cite{kuvcera2017vibrations}.  Outputs of the actin filament processors can be measured using controlled light waves and pulse  trains~\cite{goulielmakis2008single,baltuvska2003attosecond,nabekawa2017probing, ciappina2017attosecond}. There are ways to measure a vibration of a cell membrane, as demonstrated in \cite{jelinek2009measurement}. The vibration of the membrane might  reflect vibrations of cytoskeleton networks attached to the membrane~\cite{cifra2007electrical}, however it shows a cumulative effect of vibration of a cytoskeleton network. Due to polarity of actin units vibration modes are manifested in elector-magnetic perturbation which could be measured when existing experimental techniques will be perfected~\cite{pokorny1997vibrations,pokorny2004excitation}.

\bibliographystyle{elsarticle-num}


\end{document}